\documentclass[Journal,onecolumn]{IEEEtran}

\usepackage{fancyhdr}
\usepackage{cite}
\usepackage{caption}
\usepackage{graphicx}
\usepackage{mathtools}
\usepackage{psfrag}
\usepackage{subfigure}
\usepackage{url}
\usepackage{stfloats}
\usepackage{amsmath}
\usepackage{array}
\usepackage{fancyhdr}
\usepackage{epsfig}
\usepackage{amssymb}
\usepackage{color}
\usepackage{multirow}
\usepackage{amsmath,amssymb,amsfonts}
\usepackage{graphicx}
\usepackage{textcomp}
\usepackage{xcolor}
\usepackage[ruled,vlined]{algorithm2e}
\usepackage{algorithm2e,setspace}
\usepackage{amsmath}
\usepackage{graphicx}
\usepackage{amssymb}
\usepackage{setspace}

\usepackage{amssymb}
\usepackage{setspace}

\begin{document}
%
\title{Wireless Federated Distillation for Distributed Edge Learning with Heterogeneous Data}
%
%
%

\author{Jin-Hyun~Ahn$^{*}$,
        Osvaldo~Simeone$^{\dagger}$,
        and~Joonhyuk~Kang$^{*}$\\
        $^{*}$ Korea Advanced Institute of Science and Technology, School of Electrical Engineering, South Korea\\
        $^{\dagger}$ King’s College London, Department of Informatics, London, United Kingdom\\
        $^{*}$ wlsgus3396@kaist.ac.kr, jhkang@ee.kaist.ac.kr, $^{\dagger}$  osvaldo.simeone@kcl.ac.uk
        }

\maketitle


\begin{abstract}
Cooperative training methods for distributed machine learning typically assume noiseless and ideal communication channels. This work studies some of the opportunities and challenges arising from the presence of wireless communication links. We specifically consider wireless implementations of Federated Learning (FL) and Federated Distillation (FD), as well as of a novel Hybrid Federated Distillation (HFD) scheme. Both digital implementations based on separate source-channel coding and \textit{over-the-air computing} implementations based on joint source-channel coding are proposed and evaluated over Gaussian multiple-access channels.  
\end{abstract}


\section{Introduction}

The performance of local machine learning models implemented at mobile devices can be potentially improved via cooperative training methods that leverage communication among devices. Federated Learning (FL) is a recently proposed decentralized training technique that uses standard first-order distributed optimization updates and periodic exchanges of model parameter information between devices and a Parameter Server (PS) \cite{1}. FL has been advocated as a way to outperform standard Independent Learning (IL), which is based on separate training at each device, without requiring direct data exchange among the devices.

When implemented over capacity-constrained communication links, communication cost and latency may severely limit the performance of FL. To alleviate this problem, Federated Distillation (FD) was introduced for classification problems in \cite{2}. With FD, devices exchange the average output logit vectors, which includes one entry for each values of the classification label. Following the original work on distillation \cite{3,4,5}, logit values are used to define a regularizer for local training at each device. Even though FD generally reduces the communication cost, the accuracy gain as compared to IL are generally not as significant as for FL.
\begin{figure}[ht]
    \centering
    \includegraphics[width=60mm]{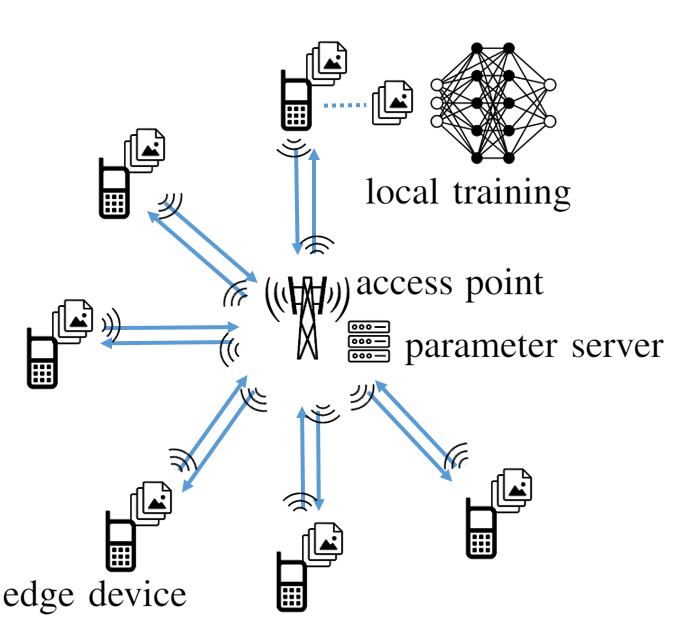}
    \caption{Edge training via wireless communications through an access point.}
    \label{fig1}
\end{figure}
 The line of work reviewed up to this point assumes noiseless and orthogonal channels between devices and PS. Wireless communication links add new challenges and opportunities due to the presence of noise and to the superposition property of wireless transmission.
 As demonstrated in \cite{Nazer}, superposition can be leveraged to enable over-the-air computing of sums of signals sent by different devices without the need to separately decode each signal. This approach has been applied to signal processing applications \cite{signalprocessing} and to transceiver beamforming design \cite{beamformingdesign}, among other problems. Recently, the idea of over-the-air computing has been leveraged to improve the efficiency of distributed gradient computation for parallelized machine learning \cite{6,7}. This is done by directly estimating the average gradient from the superposition of the signals transmitted by multiple devices over a multi-access channel, with each device transmitting its local gradient using analog communication.
 
 In light of this state of the art, in this paper, we contribute to the study of distributed learning over wireless channels by investigating the potential benefit of over-the-air computing for FL and FD. More specifically, the main contributions are as follows.  

\begin{itemize}
    \item We first propose a variant of FD that applies FD regularization to the average probability vector but also the average input per label, which is exchanged during an offline phase. This approach, referred to as Hybrid FD (HFD), can help bridge the performance gap between FL and FD;
    \item We then propose and study an implementation of FL, FD, and HFD over Gaussian multiple-access channels between devices and access point as illustrated in Fig. 1. We specifically consider both a conventional digital scheme based on quantization and channel coding, and an analog scheme that leverages over-the-air computing.
\end{itemize}

The rest of the paper is organized as follows. In Sec. \ref{PD}, we introduce the problem definition by detailing system set-up, communication models, and the baseline training protocols. In Sec. \ref{enhanced}, HFD is proposed. In Sec. \ref{sec:wireless}, both digital and analog implementation of the protocols are introduced. In Sec. \ref{NR}, the performance of the considered protocols are compared through numerical results. We conclude this paper in Sec. \ref{con}.

\section{Problem Definition}\label{PD}
\subsection{System Set-Up}
As illustrated in Fig. \ref{fig1}, we consider a wireless system including $K$ devices communicating via an Access Point (AP). Each device holds a local set $\mathbb{D}_{k}$ of data points. The goal is for each device $k$ to leverage limited communication so as to train a machine learning model that outperforms a model trained soley on the local training set $\mathbb{D}_{k}$, i.e., Independent Learning (IL). To this end, the devices communicate over a wireless shared uplink channel with the AP, which is in turn connected to a Parameter Server (PS). The protocol prescribes a number of global iterations, with each iteration $i$ encompassing \textit{local training} at each device and \textit{model information} exchange via the AP.

Focusing on a classification problem with $L$ classes, each dataset $\mathbb{D}_{k}$ consists of pairs $\left(\mathbf{c}, \mathbf{t} \right) $, where $\mathbf{c}$ is the vector of covariates and $\mathbf{t}$ is the $L\times 1$ one-hot encoding vector of the corresponding label $t\in \left\{1, \dots, L \right\}$. Each device $k\in\left\{1,\dots,K\right\}$ trains a neural network model that produces the logit vector $\mathbf{s} \left(\mathbf{c}\vert \mathbf{w}_{k} \right)$ and the corresponding output probability vector $\hat{\mathbf{t}}\left(\mathbf{c}\vert \mathbf{w}_{k} \right)$ after the last, softmax, layer, for any input $\mathbf{c}$ with the $W\times1$ weight vector $\mathbf{w}_{k}$ defining the network's operation at all layers. We define the probability vector $\hat{\mathbf{t}}$ corresponding to a logit vector $\mathbf{s}$ as
\begin{equation}
    \hat{\mathbf{t}}\left(\mathbf{s}\right)= \frac{1}{\sum\limits_{i=1}^{L} e^{s_i}}\begin{bmatrix} 
e^{s_1} \\
\vdots\\
e^{s_L}
\end{bmatrix},
\end{equation}
and we have $\hat{\mathbf{t}}\left(\mathbf{c}\vert \mathbf{w}_{k} \right)= \hat{\mathbf{t}}\left(\mathbf{s} \left(\mathbf{c}\vert \mathbf{w}_{k}\right) \right)$.

\subsection{Communication Model}
On the uplink, devices share a Gaussian multiple-access
\begin{equation}\label{receivedsignal}
\mathbf{y}=\sum_{k=1}^{K}  \mathbf{x}^{k} + \mathbf{z},
\end{equation}
where $\mathbf{x}^{k}$ is the $T\times 1 $ (real) signal transmitted by the $k$-th device, and $\mathbf{z}$ is $T\times 1 $ noise vector with independent and identically distributed (i.i.d.) $\mathcal{N} \left(0,1\right)$ entries. Each device $k$ has a power constraint $ \textrm{E}\left[ \| \mathbf{x}^{k}\|^{2}_{2}\right] /T \leq P$. Downlink broadcast communication from AP to devices is assumed to be noiseless in order to focus on the effect of the more challenging multi-access uplink channel.

\subsection{Baseline Training Protocols}\label{subsec:protocol}
In this section, we briefly review baseline training protocols, while assuming ideal noiseless communication between devices and AP. The impact of the multi-access channel will be studied in Sec. \ref{sec:wireless}.   

\textit{Independent Learning} (IL): With IL, as summarized in Algorithm 1, each learning model at device $k$ is trained on the local training set $\mathbb{D}_{k}$ by using Stochastic Gradient Descent (SGD) with step size $\alpha >0$. Training is done by minimizing the cross entropy loss. Throughout, we define the cross entropy between probability vectors $\mathbf{a}$ and $\mathbf{b}$ as $\phi(\mathbf{a},\mathbf{b})=-\sum_{l=1}^{L} a_l \log b_l$.

\textit{Federated Learning} (FL) \cite{1}: With FL, at each iteration, devices carry out a number of local updates using SGD, and then exchange the resulting weight vector with the PS. At the next iteration, the weight vectors at all devices are initialized to the average weight vector downloaded from the PS. As summarized in Algorithm 2, the algorithm can be made potentially more efficient in terms of communication by exchanging the updates of the local weight vectors rather than the vector themselves (see Sec. \ref{sec:wireless}). 

\textit{Federated Distillation} (FD) \cite{2}: Each device computes an average logit vector at the output of the softmax layer for each label $t$, where the average is computed over the SGD local iterations. Periodically, each device uploads all the $L$ averaged logit vectors to the PS. The uploaded averaged probability vectors from all devices for each label are averaged at the PS, thus obtaining $L$ global average logit vectors, one per class. Then, the devices download the global logit vectors. During local training, following distillation \cite{3}, each device is trained to optimize the weighted sum of the regular cross-entropy and of the cross-entropy between the local probability vector and the probability vector corresponding to the downloaded logit vector.

 \begin{algorithm}[H]
 \setstretch{1}
\SetAlgoLined
\SetKwBlock{Begin}{for}{}
\Begin(each device $k=1,\dots,K$){
\Begin(each iteration of local training){
\textbf{do} SGD update
 \begin{equation}\label{normalsgd}
     \mathbf{w}^{k} \leftarrow  \mathbf{w}^{k}-\alpha \nabla_{\mathbf{w}^{k}} \phi\left(\hat{\mathbf{t}} \left(\mathbf{c}\vert \mathbf{w}^{k}\right),
    \mathbf{t} \right)
 \end{equation}
for a randomly selected training example $\left(\mathbf{c},\mathbf{t}\right) \in \mathbb{D}_{k}$}}
\caption{Independent Learning (IL)}
\end{algorithm}

%

\begin{algorithm}[H]
\setstretch{1}
\SetAlgoLined
\SetKwBlock{Begin}{for}{}
\noindent\Begin(each iteration $i=1,\dots, I$){
\Begin(each device $k=1,\dots,K$){
\textbf{download} from PS the average weight update 
\begin{equation}
    \Delta\mathbf{w}_{i-1}=\frac{1}{K}\sum\limits_{k=1}^{K} \Delta\mathbf{w}_{i-1}^{k}
\end{equation}
\textbf{set} initial value $ \mathbf{w}_{i}^{k}=\mathbf{w}^{k}_{i-1}+
\Delta\mathbf{w}_{i-1}-\Delta\mathbf{w}_{i-1}^{k}  \overset{\Delta}{=} \mathbf{w}_{i,o}^{k} $\\
\SetKwBlock{Begin}{for}{end}
\Begin(each iteration of local training){\textbf{do} SGD update as in \eqref{normalsgd}, for a randomly selected training example
 $\left(\mathbf{c},\mathbf{t}\right) \in \mathbb{D}_{k}$}
\textbf{upload} update $\Delta\mathbf{w}_{i}^{k}=\mathbf{w}_{i}^{k}-\mathbf{w}_{i,o}^{k}$ to PS}}
\caption{Federated Learning (FL)}
\end{algorithm}

\begin{algorithm}[H]
\setstretch{1}
\SetAlgoLined
\SetKwBlock{Begin}{for}{}
\Begin(each iteration $i=1,\dots, I$){
\Begin(each device $k=1,\dots,K$){
\textbf{download} from PS the global-averaged logit vectors for all labels $t=1,\dots ,L$
 \begin{equation}\label{download}
\mathbf{s}_{i,t}=\frac{1}{K}\sum_{k'=1}^{K}\mathbf{s}_{i,t}^{k'}
 \end{equation}
\textbf{obtain} the local logit vectors 
 \begin{equation}\label{backslash}
\mathbf{s}_{i,t}^{\backslash k}= \frac{K \mathbf{s}_{i,t}-\mathbf{s}_{i,t}^{k} }{K-1}
\end{equation}
\textbf{initialize} $ \mathbf{s}_{i+1,t}^{ k} \coloneqq 0$ and $ n^{k}_{i+1,t}\coloneqq 0$ for all labels $t=1,\dots ,L$\\
\SetKwBlock{Begin}{for}{end}
\Begin(each iteration of local training){ 
\textbf{do} SGD update
  \begin{equation}\label{FDsgd}
     \mathbf{w}^{k}_{i} \leftarrow  \mathbf{w}^{k}_{i}-\alpha \nabla_{\mathbf{w}^{k}_{i}} \left\{\left(1-\beta \right)\phi\left(\hat{\mathbf{t}} \left(\mathbf{c}\vert \mathbf{w}_{i}^{k}\right),
    \mathbf{t} \right)+\beta \phi\left(\hat{\mathbf{t}} \left(\mathbf{c}\vert \mathbf{w}_{i}^{k}\right),
    \hat{\mathbf{t}}(\mathbf{s}_{i,t}^{\backslash k}) \right) \right\} 
    \end{equation}
for a randomly selected training example
$\left(\mathbf{c},\mathbf{t}\right) \in \mathbb{D}_{k}$\\
\textbf{update} the logit vector and the label counter
    \begin{equation}
        \mathbf{s}_{i+1,t}^{ k} \leftarrow \mathbf{s}_{i+1,t}^{ k} + \mathbf{s} \left(\mathbf{c}\vert \mathbf{w}^{k}_{i}\right)
        \end{equation}
        \begin{equation}
        n^{k}_{i+1,t} \leftarrow n^{k}_{i+1,t}+1
    \end{equation}
}
\textbf{upload} the local-averaged logit vectors $\mathbf{s}_{i+1,t}^{ k} \leftarrow \mathbf{s}_{i+1,t}^{ k}/n_{i+1,t}^{ k}$ to the PS for all labels $t=1,\dots,L$}}
\caption{Federated Distillation (FD)}
\end{algorithm}

\section{Hybrid Federated Distillation}\label{enhanced}
In this section, we propose a novel FD scheme that aims at bridging the performance gap between FD and FL at the cost of the need for an offline communication phase and of a potentially more significant leakage of
information about covariate vectors among devices. The idea is inspired by the original distillation strategy \cite{3}, for which a probability vector $\hat{\mathbf{t}}$ used during training of the distilled model for regularization is associated with the same vector of covariates $\mathbf{c}$ at both teacher and distilled models. With FD, as seen in \eqref{FDsgd}, the teacher's information $\hat{\mathbf{t}}(\mathbf{s}_{i,t}^{\backslash k})$ from other devices is associated to all covariate vectors $\mathbf{c}$ with the same label. In fact, devices only exchange the average logit vectors per label, and they do not share any covariate vector information.

The proposed Hybrid FD method (HFD) modifies FD by using not only the average logit vector exchanged at each iteration but also the average covariate vector per label, which is shared during a preliminary offline phase. Specifically, prior to the start of the global iterations, each device $k=1,\dots,K$ calculates the average covariate vectors $\tilde{\mathbf{c}}_{t}^{k}$ for the local dataset $\mathbb{D}_{k}$, which is uploaded to the PS for all labels $t=1,\dots,L$. Then, the PS calculates the global average covariate vectors for all labels $t=1,\dots,L$  
 \begin{equation}
\tilde{\mathbf{c}}_{t}=\frac{1}{K}\sum_{k'=1}^{K}\tilde{\mathbf{c}}_{t}^{k'}.
 \end{equation}
Each device $k$ downloads $\tilde{\mathbf{c}}_{t}$ and calculates the vectors
\begin{equation}
\tilde{\mathbf{c}}_{t}^{\backslash k}= \frac{K \tilde{\mathbf{c}}_{t}-\tilde{\mathbf{c}}_{t} }{K-1}
\end{equation}
for all labels $t=1,\dots,L$ in a manner similar to the logit vector \eqref{backslash}. Then, for each global iteration, each device first carries out a number of iterations during a distillation phase that operates only over the dataset of average covariate vectors. Each covariate vector is associated to the corresponding average logit vector (see \eqref{logitFD+}), and SGD is used to minimize the weighted sum of the regular cross-entropy loss and of the cross-entropy between the local probability vector and the corresponding downloaded probability vector. After the distillation phase, each device performs a number of SGD updates following the IL principle on the local dataset. The procedure hence combines both distillation and IL, and is referred to as Hybrid FD (HFD). The full algorithm is summarized in Algorithm 4.

\begin{algorithm}[H]
\setstretch{1}
\SetAlgoLined
\SetKwBlock{Begin}{for}{}
\Begin(each device $k=1,\dots,K$){
\Begin(each iteration $i=1,\dots, I$){
\textbf{download} from PS the global-averaged logit vectors \eqref{download} for all labels $t=1,\dots ,L$\\
\textbf{obtain} the logit vectors \eqref{backslash} \\
\SetKwBlock{Begin}{for}{end}
\Begin(each iteration of the distillation phase of local training){\textbf{do} SGD update as in \eqref{FDsgd} for a data point $(\tilde{\mathbf{c}}_{t}^{\backslash k}, \mathbf{t})$  for a randomly chosen label $t$}
\Begin(each iteration of the IL phase of local training){\textbf{do} SGD update as in \eqref{normalsgd} for a randomly selected training example
 $\left(\mathbf{c},\mathbf{t}\right) \in \mathbb{D}_{k}$}
\textbf{upload} the logit vectors
\begin{equation}\label{logitFD+}
    \mathbf{s}_{i+1,t}^{ k}= \mathbf{s} \left(\tilde{\mathbf{c}}_{t}^{k}\;\middle|\; \mathbf{w}^{k}_{i}\right)
\end{equation}
to the PS for all labels $t=1,\dots,L$}}
\caption{Hybrid Federated Distillation (HFD)}
\end{algorithm}

\section{Wireless Cooperative Training}\label{sec:wireless}

In this section, we consider wireless implementations of the cooperative training schemes summarized in Sec. \ref{subsec:protocol} and \ref{enhanced}. We specifically develop both a standard digital implementations based on separate source-channel coding and analog implementations based on over-the-air computing and joint source-channel coding.

Before describing both implementations for FL, FD, and HFD, it is useful to define the following functions. The function $\mathrm{sparse}_{q} \left(\mathbf{u}\right)$ introduced in \cite{8} operates as follows. First, all elements of the input vector $\mathbf{u}$ are set to zero except for the largest $q$ elements and the smallest $q$ elements. The mean values of the remaining positive elements and negative elements are respectively denoted by $\mu^{+}$ and $\mu^{-}$. If $\mu^{+}>\left|\mu^{-} \right| $, the negative elements are then set to zero and all the elements with positive values are set to $\mu^{+}$, and vice versa $\left|\mu^{-} \right| > \mu^{+} $. Furthermore, the function $\mathrm{thresh}_{q} \left(\mathbf{u}\right)$ sets all elements of the input vector $\mathbf{u}$ to zero except the $q$ elements with the largest absolute values. Finally, function $Q_{b} \left( \mathbf{u} \right)$ quantizes each non-zero element of input vector $\mathbf{u}$ using a uniform quantizer with $b$ bits per each non-zero element.

\subsection{Digital Transmission}
Under a conventional digital implementation, all devices share equally the uplink capacity of the channel \eqref{receivedsignal}, so that the number of bits that can be transmitted from each device per global iteration is given, using Shannon's capacity, as \cite{Thomas} 
\begin{equation}
    B_{D} = \frac{T}{2K} \log_2 \left(1+ K P \right). 
\end{equation}
In order to enable transmission of the analog vectors required by FL, FD, and HFD, each device compresses the corresponding information to be sent to the AP to no more than $B_{D}$ bits. Details for each scheme are provided next.

\subsubsection{FL}
Under FL, as seen in Algorithm 2, each device $k$ at the $i$-th global iteration sends the $W \times 1$ vector
$\Delta \mathbf{w}_{i}^{k}$ to the AP. To this end, we adopt sparse binary compression with error accumulation  
\cite{7,8}. Accordingly, each device $k$ at the $i$-th global iteration computes the vector $\mathbf{v}_{i}^{k}=\mathrm{sparse}_{q} \left( \Delta \mathbf{w}_{i}^{k} +\Delta_{i}^{k}\right)$, where the accumulated quantization error is updated as \cite{7,8} 
\begin{equation}\label{errorupdate}
    \Delta_{i}^{k+1}=\Delta \mathbf{w}_{i}^{k}+\Delta_{i}^{k}- Q_{b} \left(\mathbf{v}_{i}^{k}\right).
\end{equation}
Then it sends the $b$ bits obtained through the operation $Q_{b} \left(\mu \right)$, where $\mu$ is the non-zero element of $\mathbf{v}_{i}^{k}$, along with $\log_2 \binom{W}{q}$ bits specifying the indices of the $q$ non-zero elements in $\mathbf{v}_{i}^{k}$. The total number of bit to be sent by each device is hence given as 
\begin{equation}
       B_{FL} = b+ \log_2 \binom{W}{q},
\end{equation}
where $q$ is chosen as the largest integer satisfying $B_{FL} \leq B_D$ for a given bit resolution $b$.

\subsubsection{FD and HFD}

Under FD and HFD, each device $k$ at the $i$-th global iteration should send the $L \times 1$ logit vector $\mathbf{s}_{i+1,t}^{k}$ for all labels $t\in \left\{ 1,\dots, L \right\}$. To this end, each device computes the vector
$\mathbf{q}_{i,t}=Q_{b} \left(\mathrm{thresh}_{q} \left(\mathbf{s}_{i+1,t}^{k}\right)\right)$, and the resulting bits are sent to the PS, along with the positions of the non-zero entries in vector $\mathbf{q}_{i,t}$. The number of bits to be sent is hence given as 
\begin{equation}
    B_{FD}=L\left( bq+ \log_2 \binom{L}{q}\right),
\end{equation}
where $q$ is chosen the largest integer satisfying $B_{FD} \leq B_D$. 




\subsection{Analog Transmission for Over-the-Air Computing}
Under over-the-air computing, all the devices transmit their information simultaneously in an uncoded manner to the AP, and the AP decodes directly the desired sum from the received signal \eqref{receivedsignal}. Details for each scheme are provided below.

\subsubsection{FL}

In order to enable dimensionality reduction, a pseudo-random matrix $\mathbf{A} \in \mathbb{R}^{ T \times W}$ with i.i.d. entries $N(0,1/T)$ is generated and shared between the PS and the devices before the start of the protocol. 
In a manner similar to \cite{7}, each device $k$ at the $i$-th global iteration computes the vector $\mathbf{v}_{i}^{k}=\mathrm{thresh}_{q} \left( \Delta \mathbf{w}_{i}^{k} +\Delta_{i}^{k}\right)
$, where $\Delta_{i}^{k}$ denotes the accumulated error, which is updated as \eqref{errorupdate}. Then, each device $k$ transmits the vector $\mathbf{x}_i= \gamma \mathbf{A}\mathbf{v}_{i}^{k} \in \mathbb{R}^{T \times 1}$ where the scalar factor $\gamma$ is chosen to satisfy the power constraints of devices as 
\begin{equation}
    \max\limits_{k} \frac{ \gamma^2 \|\mathbf{A}\mathbf{v}_{i}^{k} \|_{2}^{2}}{T} \leq P.
\end{equation}
Finally, the PS estimates the vector $\sum_{k'=1}^{K} \mathbf{v}_{i}^{k'}$
by applying the approximate message passing (AMP) algorithm \cite{AMP} to the received signal \eqref{receivedsignal}.

\subsubsection{FD and HFD}
Under FD and HFD, each device $k$ at the $i$-th global iteration should communicate the $L \times 1$ logit vector $\mathbf{s}_{i+1,t}^{k}$ for all labels $t\in \left\{ 1,\dots, L \right\}$. We assume here that the number $T$ of channel uses is greater $L^2$, since $L$ is typically small. Otherwise, a compression scheme as described above could be readily used. We specifically assume the condition $\rho L^2 \leq T$ for source integer bandwidth expansion factor $\rho \geq 1$. Under this condition, each device $k$ at the $i$-th global iteration transmits 
\begin{equation}\label{powernormal}
    \sqrt{\frac{T P}{\max\limits_{k} \|\mathbf{R}_{\rho}\mathbf{s}_{i+1}^{k} \|_{2}^{2}}} \mathbf{R}_{\rho}\mathbf{s}_{i+1}^{k},
\end{equation}
where matrix $\mathbf{R}_{\rho}=\mathbf{1}_{\rho} \otimes \mathbf{I}_{L^2}$, with $\mathbf{1}_{\rho}=(1,\dots,1)^{T}$, implements repetition coding with redundancy $\rho$; $\mathbf{I}_{L^2}$ is a $L^2 \times L^2$ identity matrix; and we have $\mathbf{s}_{i+1}^{k} = \left[\left(\mathbf{s}_{i+1,1}^{k}\right)^{T} , \dots, \left(\mathbf{s}_{i+1,L}^{k}\right)^{T} \right]^{T}$. Note that average power constraint is satisfied thanks to the normalization in \eqref{powernormal}. 

\section{Numerical results}\label{NR}
In this example, devices train a Convolutional Neural Network (CNN) to carry out image classification based on subsets of the MNIST dataset available at each device. As in \cite{2}, we randomly select disjoint sets of $1000$ samples from the training MNIST examples, and allocate each set to a device. Each dataset is partitioned into $10$ subsets according to the ground-truth labels. Then, for each device, we select three different ``target'' labels, namely labels $3$, $6$, $9$ for device $1$, labels $2$, $5$, $8$ for device $2$, and $1$, $4$, $7$ for device $3$. We finally eliminate all the samples of the target labels except for $5$ randomly chosen samples per label.

Each device trains a $6$-layer CNN that consists of $2$ convolutional layers, $2$ max-pooling layer, $2$ fully-connected layers, and softmax layer. The devices conduct local training with batch size $64$. Each global iteration consists of $3520$ local iterations, with up to $10$ global iterations under FL and FD. For HFD, we separate $3520$ local iterations into $1408$ iterations for the distillation phase and $2112$ for the IL phase. For fair comparison, under IL, each device implements $35200$ local iterations. In Table 1, all the system parameters are summarized.

\begin{table}[ht]
\centering
\caption{Parameters}
\label{tab}
\begin{tabular}{|l|c|}
\hline
Learning rate, $\alpha$ & $0.001$  \\
  \hline
Number of weights, $W$ & $26722$ \\
  \hline
Coefficient for weighted average of cross-entropy, $\beta$
  & $0.01$\\
  \hline
Number of global iterations, $I$ & $10$\\
  \hline
Number of devices, $K$ & $3$\\
 \hline
Number of labels, $L$ & $10$\\
 \hline
Number of target labels & $3$\\
\hline
Number of bits for quantization, $b$ & $16$  \\
\hline
 Threshold level $q$ of $\textrm{thresh}_{q}$ with analog transmission of FL& $T/20$ \\
\hline
\end{tabular}
\end{table}

For reference, we first compare the performance of the training protocols in the ideal case of noiseless and orthogonal communication links to the PS. The performance metric is defined as the test accuracy measured over $10000$ randomly selected images from the MNIST dataset. In Fig. 2, we plot the test accuracy as a function of the label $t$ for device 1. It is seen that the accuracy for target labels $3$, $6$, and $9$ are significantly affected by the choice of the training protocols. We also observe that the HFD can partially bridge the performance gap between FL and FD. The accuracy of the target labels for all devices is summarized in Table II.
\begin{figure}[ht]
        \centering
    \includegraphics[width=100mm]{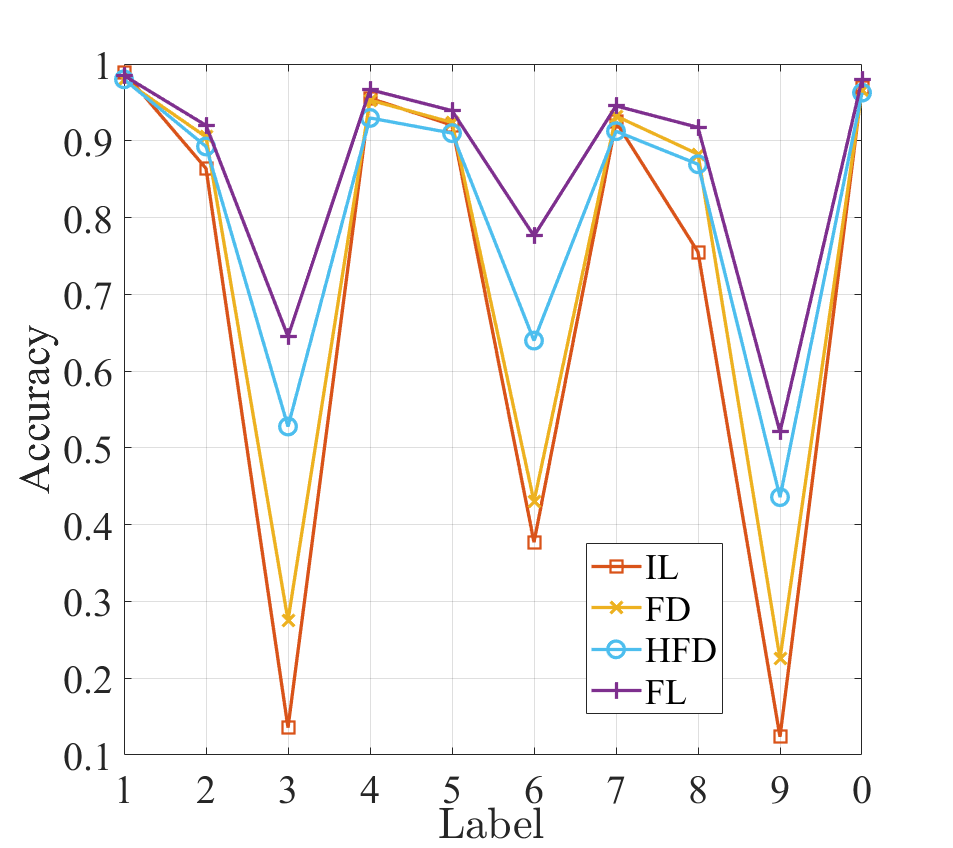}
    \caption{Classification accuracy for Device $1$ for all ten labels under IL, FL, FD, and HFD with ideal communication links.}
    \label{fig2}
\end{figure}

\begin{table}[ht]
\centering
\caption{Accuracy for target labels under IL, FL, FD, and HFD with ideal communication links.}
\label{Nochannel}
\begin{tabular}{|c|c|c|c|c|}
\hline
\multirow{2}{*}{} & \multirow{2}{*}{Device 1} & \multirow{2}{*}{Device 2} & \multirow{2}{*}{Device 3} & \multirow{2}{*}{\textbf{Average}} \\
                  &                           &                           &                           &                                   \\ \hline
IL                & 0.2122                    & 0.2132                    & 0.3758                    & \textbf{0.2671}                   \\ \hline
FD                & 0.3103                    & 0.2581                    & 0.4238                    & \textbf{0.3307}                   \\ \hline
HFD               & 0.5345                    & 0.4649                    & 0.6004                    & \textbf{0.5333}                   \\ \hline
FL                & 0.6472                    & 0.6197                    & 0.7410                    & \textbf{0.6693}                   \\ \hline
\end{tabular}
\end{table}

\newpage
\begin{figure}[ht]
    \centering
    \includegraphics[width=100mm]{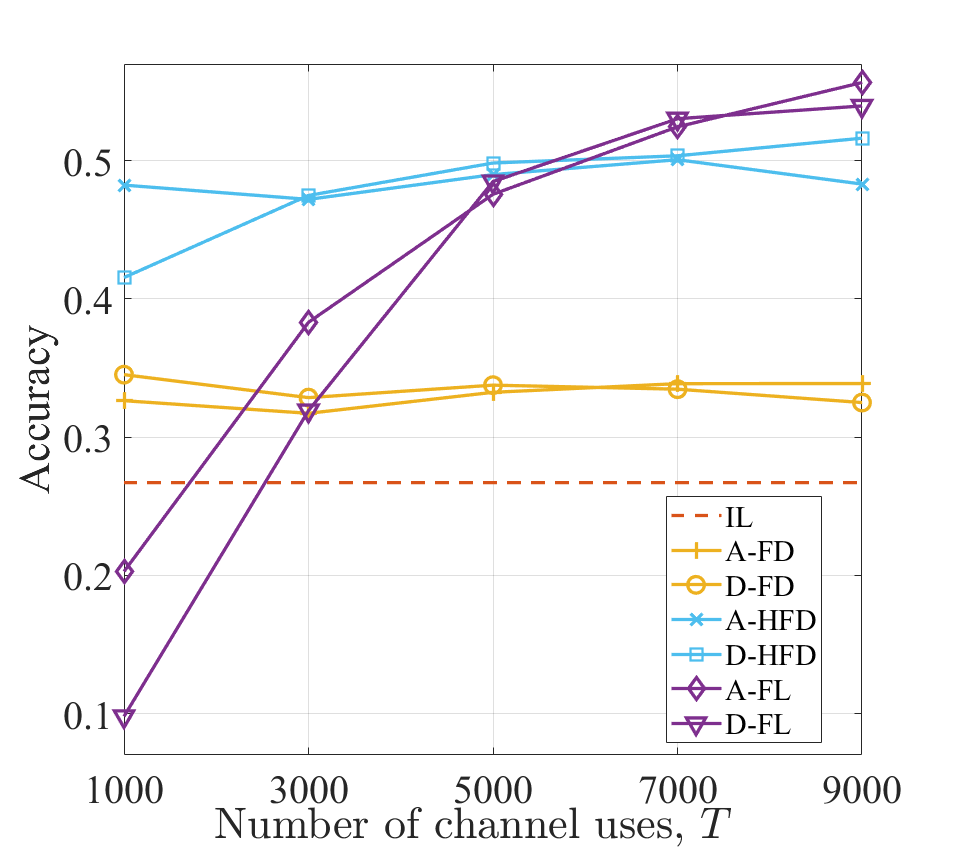}
    \caption{Classification accuracy under IL, FL, FD, and HFD for digital and analog implementations.}
    \label{fig3}
\end{figure}

\begin{figure}[ht]
    \centering
    \includegraphics[width=100mm]{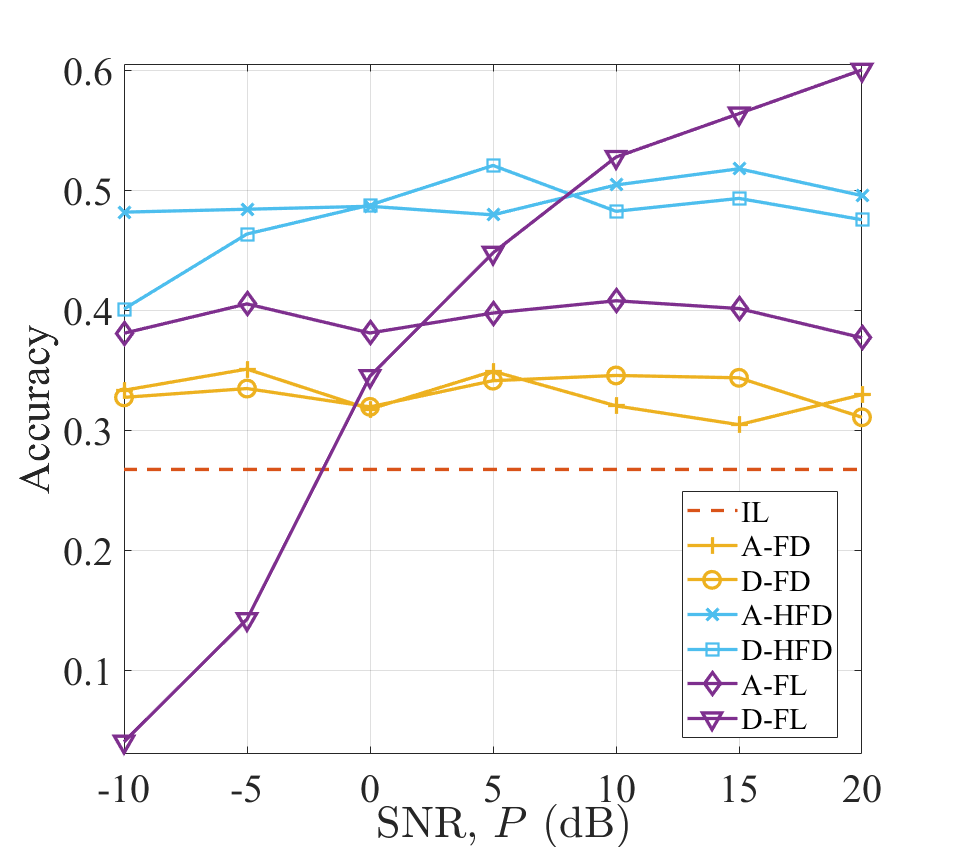}
    \caption{Classification accuracy under IL, FL, FD, and HFD for digital and analog implementations.}
    \label{fig4}
\end{figure}
In Fig. 3 and Fig. 4, the performance of FL, FD, and HFD is shown when the the protocols are implemented via the digital and analog transmission schemes introduced in Sec. \ref{sec:wireless}. We denote the digital transmission schemes by prefix ``D-'' and the analog transmission for over-the-air computing by the prefix ``A-''. In Fig. 3, the number $T$ of channel uses increases from $1000$ to $9000$ and the signal-to-noise ratio (SNR) is $P=0$ dB. It is first observed that the implementations of FD, HFD are more robust at low values of $T$ than FL, which has a larger communication overhead. In fact, FL can even be outperformed by IL for sufficiently low values of $T$. Furthermore, HFD is seen to uniformly outperform FD and all other schemes, unless $T$ is sufficiently large, in which case FL is to be preferred. Finally, analog and digital implementations are seen to perform comparably in this regime.

In Fig. 4, the SNR $P$ increases from $-10$dB to $20$dB and the number of channel uses is $T=3000$. First, it is observed that, thanks to their low communication overhead, both analog and digital implementations of FD and HFD perform well also in the low-SNR regime. The same is true for the analog implementation of FL, which can benefit from analog compression. In contrast, D-FD requires a larger SNR in order to obtain a sufficiently accurate data transmission, in which case it is able to outperform all schemes.

\section{Conclusions}\label{con}
In this work, we have considered wireless implementations of Federated Learning (FL), Federated Distillation (FD), and of a novel Hybird Federated Distillation (HFD) scheme, under both digital and analog implementations over Gaussian multiple-access channels. While FL is the best strategy in the presence of ideal communication channels, FD and HFD have significantly lower communication overhead than FL. Accordingly, it was seen via numerical results that both digital and analog implementations of FD and HFD can outperform FL whenever the communication links are limited in terms of number of channel uses (bandwidth) or SNR. Furthermore, the proposed HFD can significantly improve the accuracy obtained via FD.

\section*{Acknowledgments}
The work of J. Ahn and J. Kang was supported by the National Research Foundation of Korea (NRF) grant funded by the Korea government (MSIT) (No. 2017R1A2B2012698). The work of O. Simeone was supported by the European Research Council (ERC) under the European Union's Horizon 2020 research and innovation programme (grant agreement No. 725731).

\ifCLASSOPTIONcaptionsoff
  \newpage
\fi



%

\end{document}